\documentclass[journal]{IEEEtran}
\usepackage{cite,units}
\ifCLASSINFOpdf
   \usepackage[pdftex]{graphicx}
  \DeclareGraphicsExtensions{.pdf,.jpeg,.png,.eps}
\else
  \usepackage[dvips]{graphicx}
   \graphicspath{{../eps/}}
  \DeclareGraphicsExtensions{.eps}
\fi
\usepackage[tight,footnotesize]{subfigure}
\usepackage[cmex10]{amsmath}
\usepackage{cleveref}
\usepackage{mdwmath}
\usepackage{mdwtab}

\crefname{equation}{equation}{equations}
\Crefname{equation}{Equation}{Equations}
\crefrangelabelformat{equation}{(#3#1#4--#5#2#6)}
\crefmultiformat{equation}{equations (#2#1#3}{, #2#1#3)}{#2#1#3}{#2#1#3}
\Crefmultiformat{equation}{Equations (#2#1#3}{, #2#1#3)}{#2#1#3}{#2#1#3}
\allowdisplaybreaks[2]
\begin{document}
\title{Below-Cutoff Propagation in Metamaterial-Lined Circular Waveguides}
\author{Justin~G.~Pollock,~\IEEEmembership{Student~Member,~IEEE,}
        and~Ashwin~K.~Iyer,~\IEEEmembership{Member,~IEEE,}%
}
\markboth{}%
{Shell \MakeLowercase{\textit{et al.}}: IEEE Journals}
\maketitle
\begin{abstract}
This paper investigates the propagation characteristics of circular waveguides whose interior surface is coated with a thin metamaterial liner possessing dispersive, negative, and near-zero permittivity. A field analysis of this system produces the dispersion of complex modes, and reveals in detail intriguing phenomena such as backward-wave propagation below the unlined waveguide's fundamental-mode cutoff, resonant tunneling of power, field collimation, and miniaturization. It is shown how the waveguide geometry and metamaterial parameters may be selected to engineer the lined waveguide's spectral response. Theoretical dispersion and transmission results are closely validated by full-wave simulations.
\end{abstract}

\section{Introduction}
\label{sec:intro}

\IEEEPARstart{H}OLLOW waveguides are widely used in applications requiring high power-handling capability and simple integration with radiating devices such as horn antennas. As such, they are ideally suited to microwave radio links, radar, and satellite communications. It is well known that air-filled waveguides enclosed by perfect-electric-conducting (PEC) boundaries support a discrete spectrum of modes, each possessing a cutoff frequency, $f_c$, above which it is capable of propagating power. Homogeneously filling the vacuum region of the waveguide with an isotropic dielectric material serves to reduce these cutoff frequencies, without modifying the corresponding transverse modal field distributions. However, it is sometimes desirable to have access to the enclosed region of a waveguide, making it convenient to partially fill it with a dielectric to achieve similar results. These inhomogeneous structures have special impedance-boundary conditions and support modal field distributions that can differ significantly from those of their conventional, homogeneously-filled counterparts.

One such instance is the PEC circular waveguide containing two concentric dielectric regions. This structure has been extensively investigated in the form of a waveguide loaded with a dielectric rod~\cite{KWwhite,clarricoats1964evanescent}. However, the same analysis can be applied to its variant: a waveguide whose interior wall is lined with a dielectric medium. This structure supports hybrid electric ({\em HE}) and hybrid magnetic ({\em EH}) modes which are similar to the transverse electric ({\em TE}) and transverse magnetic ({\em TM}) modes of a homogeneously filled circular waveguide, except that the longitudinal electric and magnetic fields do not, in general, vanish. In certain cases, such as circularly symmetric modes or near cutoff, the dielectric-lined circular waveguide's {\em HE} and {\em EH} modes can be likened to the {\em TE} and {\em TM} modes, respectively, of its homogeneously-filled counterpart. The inhomogeneous nature of the waveguide volume introduces additional degrees of freedom in engineering its modal dispersion characteristics, such as the dielectric layer's permittivity, permeability, and thickness, which may be used to enhance bandwidth and introduce backward-wave propagation~\cite{tsandoulas1973bandwidth,clarricoats1964evanescent}.

Metamaterial loading of waveguides is a new avenue of research in which the propagation characteristics of an unloaded waveguide are altered through the inclusion of metamaterials, whose characteristics may be engineered to achieve intriguing propagation phenomena. For example, the theory of images enables propagation inside an infinite array of metamaterial inclusions to be described by placing a finite number of such inclusions inside a waveguide. In fact, it has been shown that propagation can be restored in below-cutoff waveguides by loading them with chains of magnetic and/or electric scatterers, such as the split-ring resonator (SRR), complimentary split-ring-resonator (CSRR), or wire-lines~\cite{hrabar2005waveguide,meng2011controllable}. In these works, the metamaterial-loaded waveguides are often treated as homogeneously filled by a material possessing epsilon-positive (EPS), mu-positive (MPS), epsilon-negative (ENG), and/or mu-negative (MNG) responses. Simultaneous EPS and MPS is referred to as `double-positive' (DPS), and simultaneous ENG and MNG is referred to as `double-negative' (DNG). However, whereas homogeneously filled models may explain certain elementary dispersion features, they cannot describe the full dispersion of complex modes, the distribution of power in the transverse cross-section, and the possibility of backward coupling of power, all of which have been observed in inhomogeneously filled structures. Furthermore, these homogeneous models prove inadequate in applications in which the waveguide interior must remain empty. Therefore, a more complete description of propagation phenomena in metamaterial-loaded waveguides demands an inhomogeneous model.

In 2004, Al\`{u} et al observed that the propagation of {\em TE} and {\em TM} modes can be made independent of the thickness of a parallel-plate waveguide, provided that it is filled inhomogeneously with bilayers of the varieties DPS-DNG, DNG-ENG, DPS-MNG and ENG-MNG~\cite{alu2004guided}. It was later shown by Silveirinha and Engheta that complete transmission or `supercoupling' could be achieved in discontinuous waveguides of arbitrarily small cross section, provided that they are homogeneously filled with single-negative (SNG) metamaterials possessing epsilon-near-zero (ENZ) values~\cite{enghati_tunnel}. Subsequent experimental validations of supercoupling employed either a waveguide containing an embedded metamaterial with zero permittivity, such as the CSRR~\cite{liu_experiment_tunnel}, or a narrow waveguide channel that naturally exhibits zero effective permittivity at its cutoff frequency~\cite{alu2010coaxial}.

Whereas the standard analysis of the dielectric-lined circular waveguide has been applied to the case of metamaterial liners~\cite{rao2007propagation,brand2009dispersion}, these works have either omitted the necessary frequency dispersion, concerned themselves with the DNG case alone, or limited their study to obtaining the field expressions and dispersion relations. The present work investigates the intriguing effects and implications of partially filling circular waveguides with thin, dispersive ENZ metamaterial liners and establishes several phenomena, including below-cutoff propagation, resonant tunneling and backward coupling of power, field collimation, and the possibility of waveguide miniaturization. Section~\ref{sec:theory} presents a rigorous hybrid-mode analysis to determine the lined waveguide's general dispersion features, cutoff frequencies, and field configurations. Section~\ref{sec:design} presents a representative metamaterial-lined waveguide design, analyzes its dispersion features, and offers design guidelines for engineering the frequency response of its {\em HE}$_{11}$ mode. Finally, Sec.~\ref{sec:trans} presents full-wave dispersion and transmission simulations showing close agreement with the theoretical results. Potential applications of this study include the design of metamaterial-lined waveguides to demonstrate reverse Cherenkov radiation, as in Ref.~\cite{duan2009research}, and the inclusion of metamaterials inside magnetic-resonance scanners to enable traveling-wave imaging at low static field strengths, as proposed in Ref.~\cite{PollockISMRM2012}.
\section{Theory}
\label{sec:theory}

\begin{figure}[!t]
\centering
\includegraphics[width=1.5in]{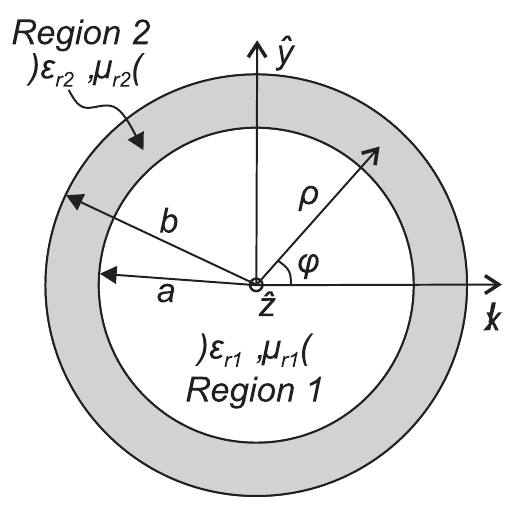}
\caption{\unskip Transverse cross-section of the dielectric-lined circular waveguide consisting of two concentric dielectric regions. An inner core region of radius $a$ and a liner of thickness $b-a$ are filled with materials described by relative parameters \{$\epsilon_{r1}$, $\mu_{r1}$\} and \{$\epsilon_{r2}$, $\mu_{r2}$\}, respectively.}
\label{fig1}
\end{figure}

Figure~\ref{fig1} presents the geometry of the dielectric-lined circular waveguide under consideration. An inner core region of permittivity $\epsilon_1=\epsilon_{r1}\epsilon_0$ and permeability $\mu_1=\mu_{r1}\mu_0$ is surrounded by a dielectric material of thickness $t=b-a$ with permittivity $\epsilon_2=\epsilon_{r2}\epsilon_0$ and permeability $\mu_2=\mu_{r2}\mu_0$. The dielectric-lined waveguide modes are no longer pure {\em TE}$_{mn}$ and {\em TM}$_{mn}$ modes; they are, instead, classified as {\em HE}$_{mn}$ and {\em EH}$_{mn}$ modes generally possessing both $H_z\neq0$ and $E_z\neq0$. We now present the analysis of the dielectric-lined circular waveguide, which may also be found in several prior works~\cite{KWwhite,clarricoats1964evanescent,rao2007propagation}, in order to provide the reader with some context for the main analytical results of this work. Suppressing the propagation term ($e^{-\gamma z}$) and the time-harmonic nature of the fields ($e^{j\omega t}$), the electric- and magnetic-field components in the axial ($z$-) direction in region 1 ($0\leq\rho\leq a$) may be represented by
\begin{subequations}
\begin{align}
E_{1z}&=C_1J_n(k_{1\rho}\rho)\cos(n\phi)\label{EQ1a}\\[3mm]
H_{1z}&=C_2J_n(k_{1\rho}\rho)\sin(n\phi)\label{EQ1b}
\end{align}
\end{subequations}
\noindent where $C_1$ and $C_2$ are the (generally complex) amplitude coefficients, $J_n(k_{1\rho}\rho)$ is a Bessel function of the first kind, $n$ is the azimuthal mode index, $k^2_{1\rho}=k_1^2-\gamma^2$, $k_1=w(\epsilon_1\mu_1)^{0.5}$, and $\gamma$ is the axial propagation constant. In region 2, the radially outgoing and incoming fields are represented by Bessel and Neumann functions, $J_n(k_{2\rho}\rho)$ and $Y_n(k_{2\rho}\rho)$. Since the tangential electric-field components must vanish at the PEC boundary $(\rho=b)$, the solution in region 2 ($a\leq\rho\leq b$) can be shown to have the following form:
\begin{subequations}
\begin{align}
E_{2z}&=C_3F_n(k_{2\rho}\rho)\cos(n\phi)\label{EQ2a}\\[3mm]
H_{2z}&=C_4G_n(k_{2\rho}\rho)\sin(n\phi)\label{EQ2b}
\end{align}
\end{subequations}
where
\begin{subequations}
\begin{align}
F_n(k_{2\rho}\rho)&=Y_n(k_{2\rho}b)J_n(k_{2\rho}\rho)-J_n(k_{2\rho}b)Y_n(k_{2\rho}\rho)\label{EQ3a}\\[3mm]
G_n(k_{2\rho}\rho)&=Y'_n(k_{2\rho}b)J_n(k_{2\rho}\rho)-J'_n(k_{2\rho}b)Y_n(k_{2\rho}\rho)\label{EQb}
\end{align}
\end{subequations}
\noindent Here, $C_3$ and $C_4$ are the amplitude coefficients, $k^2_{2\rho}=k_2^2-\gamma^2$, and $k_2=w(\epsilon_2\mu_2)^{0.5}$. Knowing the longitudinal field components, the tangential field components can be obtained using the transverse decomposition of Maxwell's equations. The transverse components of the electric and magnetic fields (that is, the $\rho$ and $\phi$ components) in both regions are given by
\begin{subequations}
\begin{align}
E_{1\rho}&=(\frac{-\gamma}{k_{1\rho}}C_1J'_n(k_{1\rho}\rho)-\frac{j\omega\mu_1 n}{k^2_{1\rho}\rho}C_2J_n(k_{1\rho}\rho))\cos(n\phi)\label{EQ4a}\\
E_{1\phi}&=(\frac{\gamma n}{k^2_{1\rho}\rho}C_1J_n(k_{1\rho}\rho)+\frac{j\omega\mu_1}{k_{1\rho}}C_2J'_n(k_{1\rho}\rho))\sin(n\phi)\label{EQ4b}\\
E_{2\rho}&=(\frac{-\gamma}{k_{2\rho}}C_3F'_n(k_{2\rho}\rho)-\frac{j\omega\mu_2 n}{k^2_{2\rho}\rho}C_4G_n(k_{2\rho}\rho))\cos(n\phi)\label{EQ4c}\\
E_{2\phi}&=(\frac{\gamma n}{k^2_{2\rho}\rho}C_3F_n(k_{2\rho}\rho)+\frac{j\omega\mu_2}{k_{2\rho}}C_4G'_n(k_{2\rho}\rho))\sin(n\phi)\label{EQ4d}\\
H_{1\rho}&=(\frac{-\gamma}{k_{1\rho}}C_2J'_n(k_{1\rho}\rho)-\frac{j\omega\epsilon_1 n}{k^2_{1\rho}\rho}C_1J_n(k_{1\rho}\rho))\sin(n\phi)\label{EQ4e}\\
H_{1\phi}&=(\frac{-\gamma n}{k^2_{1\rho}\rho}C_2J_n(k_{1\rho}\rho)-\frac{j\omega\epsilon_1}{k_{1\rho}}C_1J'_n(k_{1\rho}\rho))\cos(n\phi)\label{EQ4f}\\
H_{2\rho}&=(\frac{-\gamma}{k_{2\rho}}C_4G'_n(k_{2\rho}\rho)-\frac{j\omega\epsilon_2 n}{k^2_{2\rho}\rho}C_3F_n(k_{2\rho}\rho))\sin(n\phi)\label{EQ4g}\\
H_{2\phi}&=(\frac{-\gamma n}{k^2_{2\rho}\rho}C_4G_n(k_{2\rho}\rho)-\frac{j\omega\epsilon_2}{k_{2\rho}}C_3F'_n(k_{2\rho}\rho))\cos(n\phi)\label{EQ4h}
\end{align}
\end{subequations}
where
\begin{subequations}
\begin{align}
F'_n(k_{2\rho}\rho)&=Y_n(k_{2\rho}b)J'_n(k_{2\rho}\rho)-J_n(k_{2\rho}b)Y'_n(k_{2\rho}\rho)\label{EQ5a}\\[3mm]
G'_n(k_{2\rho}\rho)&=Y'_n(k_{2\rho}b)J'_n(k_{2\rho}\rho)-J'_n(k_{2\rho}b)Y'_n(k_{2\rho}\rho)\label{EQ5b}
\end{align}
\end{subequations}
The continuity of the tangential electric- and magnetic-field components at $\rho=a$ relate the coefficients to one another as follows:
\begin{subequations}
\begin{align}
C_3&=C_1\frac{J_n(k_{1\rho}a)}{F_n(k_{2\rho}a)},\;\;C_4=C_2\frac{J_n(k_{1\rho}a)}{G_n(k_{2\rho}a)}\label{EQ6a}\\[3mm]
\frac{C_1}{C_2}&=\frac{ \frac{\gamma n}{j\omega a}[\frac{1}{k^2_{2\rho}} - \frac{1}{k^2_{1\rho}}]}{\frac{\epsilon_1}{k_{1\rho}}\frac{J'_n(k_{1\rho}a)}{J_n(k_{1\rho}a)}-\frac{\epsilon_2}{k_{2\rho}}\frac{F'_n(k_{2\rho}a)}{F_n(k_{2\rho}a)}}\label{EQ6b}\\[3mm]
\frac{C_2}{C_1}&=\frac{ \frac{\gamma n}{j\omega a}[\frac{1}{k^2_{2\rho}} - \frac{1}{k^2_{1\rho}}]}{\frac{\mu_1}{k_{1\rho}}\frac{J'_n(k_{1\rho}a)}{J_n(k_{1\rho}a)}-\frac{\mu_2}{k_{2\rho}}\frac{G'_n(k_{2\rho}a)}{G_n(k_{2\rho}a)}}\label{EQ6c}
\end{align}
\end{subequations}
\noindent where equation (6b) is used for {\em EH} modes and equation (6c) is used for {\em HE} modes.
Using equations (6b) and (6c), the coefficients $C_1$ and $C_2$ can be eliminated to obtain the following dispersion relation:
\begin{subequations}
\begin{align}
A\cdot B&=(\frac{\gamma n}{\omega a})^2 [\frac{1}{k^2_{2\rho}} - \frac{1}{k^2_{1\rho}}]^2\label{EQ7c}\\
\intertext{where}
A&=[\frac{\epsilon_1}{k_{1\rho}}\frac{J'_n(k_{1\rho}a)}{J_n(k_{1\rho}a)}-\frac{\epsilon_2}{k_{2\rho}}\frac{F'_n(k_{2\rho}a)}{F_n(k_{2\rho}a)}]\label{EQ7a}\\[3mm]
B&=[\frac{\mu_1}{k_{1\rho}}\frac{J'_n(k_{1\rho}a)}{J_n(k_{1\rho}a)}-\frac{\mu_2}{k_{2\rho}}\frac{G'_n(k_{2\rho}a)}{G_n(k_{2\rho}a)}]\label{EQ7b}
\end{align}
\end{subequations}
Dielectric-lined waveguides are inhomogeneous structures that support hybrid modes with complex propagation constants of the form $\gamma=\alpha+j\beta$. Whereas nonzero $\alpha$ is attributed to dielectric or conductive loss in homogeneous waveguides, its existence in inhomogeneous waveguides is a direct result of the coupling of power between regions and can appear even in the absence of losses. Modes for which the value of $\alpha$ is substantially larger than $\beta$ are termed evanescent. These modes attenuate rapidly in the longitudinal direction, and are limited in their ability to transport power along the waveguide, much like the evanescent modes of a homogeneously filled waveguide; as a result, they shall not be the focus of this study. On the other hand, modes with $\alpha$  much smaller than $\beta$ can lead to the propagation of real power; we term these propagating modes. Modes with $\alpha$ comparable in magnitude to $\beta$ can be considered neither purely propagating nor evanescent, and are therefore termed complex modes. These complex modes exist in complex conjugate pairs, and the two conjugate modes transport power in opposite directions. Only if the conjugate modes are excited unequally can they transport real power~\cite{islam2010independence}.

Using Muller's method to determine the roots in the complex-$\gamma$ plane, it is possible to numerically arrive at a set of continuous dispersion curves. It should be noted that there exist certain limiting cases in which the modal fields revert to those of the homogeneous case. For $\gamma=0$, (i.e., at cutoff) the RHS of the dispersion relation (7a) becomes zero and the {\em HE} and {\em EH} modes are decoupled. This results in the roots of (7b) and (7c), determining, respectively, the cutoff frequencies of the {\em EH} and {\em HE} modes. Since the fundamental {\em TE}$_{11}$ mode is the most widely used in applications, this paper focuses on the propagation characteristics of the {\em HE}$_{11}$ mode.

\begin{figure}[!t]
\centering
\includegraphics[width=3in]{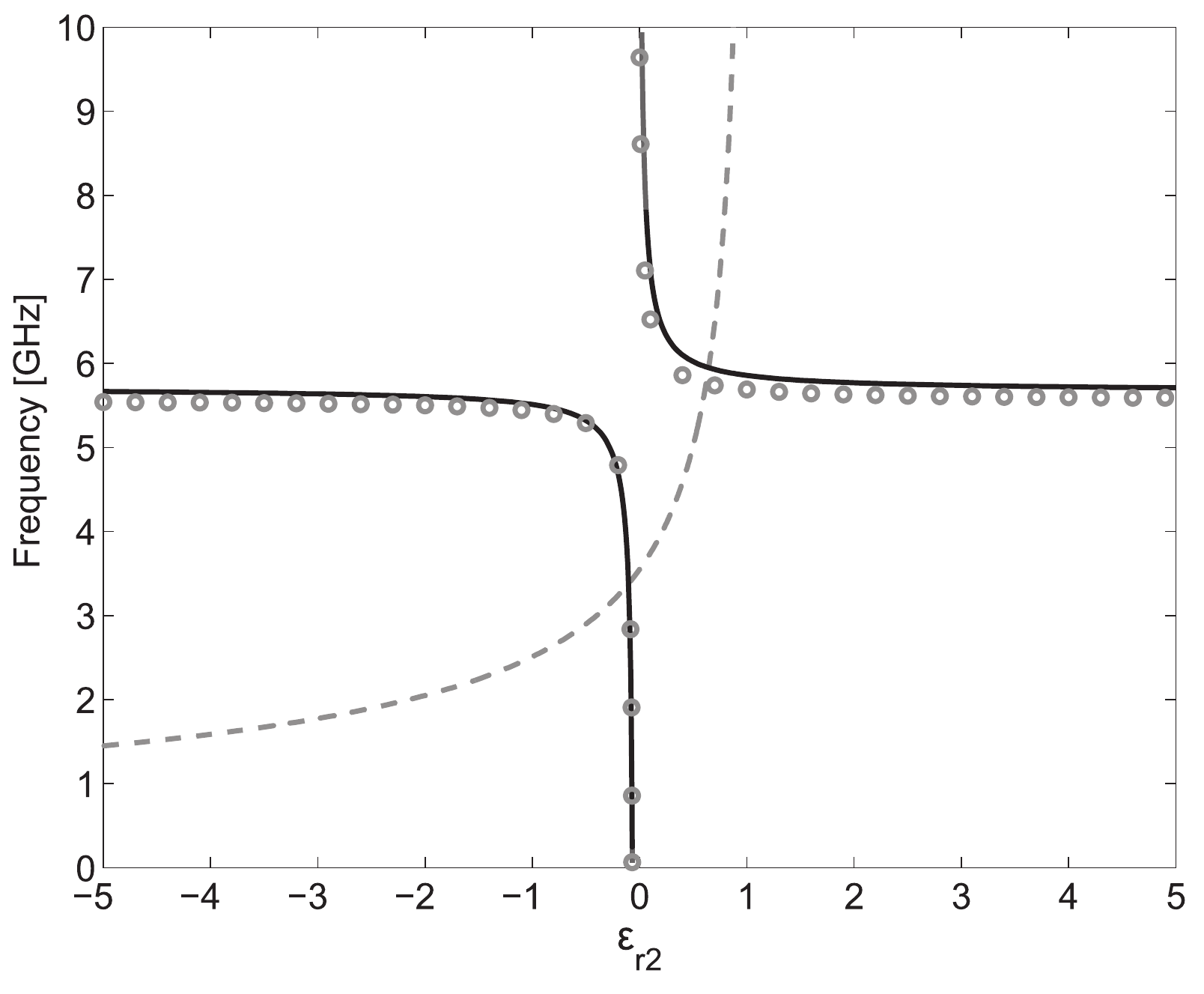}
\caption{\unskip The {\em HE}$_{11}$-mode cutoff frequency versus liner permittivity. The full dispersion relations of equation~(7a) (solid black line) is compared to the approximate expression equation~(8a) (empty grey circles). The waveguide's dimensions are $b=15$mm and $a=14$mm ($t=1$mm). The permittivity of the liner $\epsilon_{r2}(\omega)$ (dashed grey line) follows a Drude model with parameters $\omega_{ep}=3.550$GHz and $\omega_{t}=0$.}
\label{fig2:subfig1}
\end{figure}
In Fig.~\ref{fig2:subfig1}, we show the relationship between the cutoff frequency of the {\em HE}$_{11}$ mode and the liner's relative permittivity, $\epsilon_{r2}$ (solid black line), which is obtained using the analytically derived dispersion relation (7c) through enforcing the cutoff condition, $\gamma=0$. The representative circular waveguide investigated has physical dimensions $b=15$mm and $a=14$mm (liner thickness of $t=1$mm), with an inner vacuum region ($\epsilon_{r1}=\mu_{r1}=1$) and a liner region with dispersive $\epsilon_{r2}$ and a nonmagnetic response ($\mu_{r2}=1$). It should be noted that $\epsilon_{r2}$ is here an independent variable; i.e., each point on the curve is only concerned with $\epsilon_{r2}$ at a single frequency (specifically, cutoff). Figure~\ref{fig2:subfig1} may be employed in choosing the value of permittivity required to achieve a desired cutoff frequency, after which a suitable dispersive permittivity function satisfying this condition may be selected.

A well known result of the homogeneously filled dielectric waveguide is that the cutoff frequency of the {\em TE}$_{11}$ mode scales as the inverse square of permittivity, when that permittivity is assumed positive and nondispersive. Even though the metamaterial liner only occupies a small portion of the total cross-section of the waveguide, it is to be expected that a thin EPS liner with permittivity larger than unity marginally lowers the natural cutoff frequency, and conversely, that an EPS liner with permittivity smaller than unity will increase the natural cutoff frequency. Indeed, Fig.~\ref{fig2:subfig1} shows that the cutoff frequency is weakly dependent on large positive permittivity values. However, the cutoff frequency increases dramatically for small positive ENZ values, suggesting that the waveguide is thrust more deeply into cutoff as $\epsilon_{r2}\rightarrow0^+$. In this region, the liner permittivity is characterized by both EPS and ENZ, which we more compactly refer to as `epsilon positive and near zero' (EPNZ).

The cutoff frequency's dependence on negative permittivity values is even less intuitive: large, negative liner permittivities in the range shown produce marginally decreased cutoff frequencies; however, as $\epsilon_{r2}\rightarrow 0^-$, the cutoff frequency of the {\em HE}$_{11}$ mode is dramatically {\em lowered}, suggesting that a waveguide lined with a metamaterial possessing such permittivity values may support propagation at very low frequencies, well below its natural cutoff. It is evident from Fig.~\ref{fig2:subfig1} that this regime is characterized by both ENG and ENZ, which we term `epsilon negative and near zero' (ENNZ). Although not shown, it has been determined that thinner liners require more extreme ENNZ values (i.e., closer to zero) to realize the same reduction in cutoff frequency. As may be surmised from the curve, there exists an asymptote as $\epsilon_{r2}\rightarrow 0^-$ at which the cutoff frequency is reduced to dc. Of course, this is a pathological case, since it would require an ENZ material at dc and, in any event, could only result in the trivial electrostatic field solution.

Knowing that ENNZ liners reduce the cutoff frequency of the {\em HE}$_{11}$ mode, the decoupled dispersion relation for the {\em HE} modes (7c) can be simplified using small-argument approximations to reduce the Bessel and Neumann functions with the goal of obtaining a non-transcendental expression. The two conditions provided by the ENNZ liner, namely that $f_c\rightarrow 0$ and $\epsilon_{r2}\rightarrow 0^-$, validate the assumption $\{k_{2\rho}b$, $k_{1\rho}a\}\rightarrow 0$ in this frequency regime. Therefore, a first-order power-series approximation of the Bessel and Neumann functions can be applied to (7c), which yields the result that the cutoff frequency of the {\em HE}$_{11}$ mode is related to the permittivity, $\epsilon_{r2}$, of the liner and thickness, $t=b-a$, as follows:
\vspace{3pt}
\begin{subequations}
\begin{align}
f_c&=(\frac{2(2)^{0.5}c}{2\pi a})(\frac{1-K}{1-3K})^{0.5}\label{EQ8a}\\
\intertext{where}
 K&=\epsilon_{r2}\frac{1+\frac{b^2}{a^2}}{1-\frac{b^2}{a^2}}\label{EQ8b}
\end{align}
\end{subequations}
\noindent This simple relationship constitutes a design equation, since it specifies the ENNZ values of $\epsilon_{r2}$ necessary to produce a desired $f_c$, given waveguide dimensions $a$ and $b$. In Fig. \ref{fig2:subfig1}, the approximate expression (8a) (empty grey circles) is compared to the exact analytical expression (7a) (solid black line). It is evident that (8a) models the trend over all $\epsilon_{r2}$ values with good accuracy, but is expectedly most accurate in the ENNZ region in which the asymptotic limits are valid and the cutoff frequency is strongly reduced. The margin of error in this expression can be evaluated by applying the two cases describing an unloaded waveguide: one in which $\epsilon_{r2}\rightarrow 1$ and simultaneously $a\rightarrow 0$, and the other in which $a\rightarrow b$, and comparing the results to the well-known expression for the {\em TE}$_{11}$-mode cutoff frequency of a homogeneously vacuum-filled circular waveguide ($f_{c,unlined}=1.841c/(2\pi a)$). It is easily shown that the former case produces $f_c=2c/(2\pi a)$, whereas the latter case yields $f_c=1.633c/(2\pi a)$; the geometric average of these limiting cases is within $2\%$ of $f_{c,unlined}$.

Equation (8a) also seems to suggest that the cutoff frequency can be set to zero by choosing $K=1$ which, from (8b), yields the following result:
\vspace{3pt}
\begin{equation}
\epsilon_{r2,max}=\frac{1-b^2/a^2}{1+b^2/a^2}\label{EQ9}
\end{equation}
\noindent Equation (9) suggests that there exists a particular maximum ENNZ value of $\epsilon_{r2}$ that reduces the cutoff frequency to zero, which is solely dependent on the waveguide's and liner's physical dimensions. Inserting the above representative waveguide dimensions into (9), one obtains $\epsilon_{r2,max}=-0.06889$ which coincides exactly with the value of permittivity at which the asymptote occurs in Fig.~\ref{fig2:subfig1}, as expected. Thus, the liner needs only possess a {\em sufficiently} ENNZ $\epsilon_{r2}$ to effect a desired reduction in the {\em HE}$_{11}$-mode cutoff frequency. For practical metamaterial technologies, operation in this frequency-reduced regime offers the potential for miniaturization of waveguide components.

\section{Design}
\label{sec:design}
In Fig.~\ref{fig2:subfig1}, it was implicitly assumed that the liner permittivity required to produce a desired cutoff frequency could actually be achieved at that frequency. However, designing for a particular {\em HE}$_{11}$-mode cutoff requires the dispersive nature of $\epsilon_{r2}$ to be taken into account. Hence, the liner permittivity's dispersion is now described by a Drude model with $\epsilon_{r2}(\omega)=1-\omega_{ep}^2/\omega(\omega-j\omega_t)$, in which $\omega_{ep}$ is the plasma frequency and $\omega_t$ is the damping frequency establishing the liner's loss. This model accurately approximates the complex, dispersive nature of the liner's permittivity over a select frequency range. This type of dispersion can be realized by any number of existing metamaterial technologies including wire-grid media~\cite{pendry1996} or radially arranged transmission line metamaterials~\cite{pollock2011effective}, and therefore lends itself readily to practical implementation. To design the metamaterial-lined circular waveguide for a specific cutoff frequency $f_c$, the waveguide dimensions (outer radius $b$, inner radius $a$) are chosen and the corresponding required $\epsilon_{r2}$ may be determined from equation (8a). Thereafter, it remains only to choose an appropriate dispersion model for the liner permittivity such that $\epsilon_{r2}$ assumes the needed value at the desired cutoff frequency. Conversely, the dimensions of the waveguide can be determined if given an $\epsilon_{r2}(\omega)$ based on a known metamaterial frequency response.

For the chosen waveguide dimensions ($b=15$mm and $a=14$mm), it can be observed from Fig.~\ref{fig2:subfig1} that a liner permittivity of $\epsilon_{r2}=-0.09$ will result in a reduced cutoff frequency of $f_c=3.381$GHz. To achieve this goal, the Drude model parameters for the liner are set to $\omega_{ep}=3.550$GHz and $\omega_t=0$MHz. Superposing the metamaterial's dispersive permittivity (dashed grey curve) on the cutoff-frequency curve, as also shown in Fig.~\ref{fig2:subfig1}, two intersections are observed: the lower-frequency intersection corresponds to the designed {\em HE}$_{11}$-mode cutoff, which, for $\epsilon_{r2}=-0.09$, indeed produces $f_c=3.381$GHz; the higher-frequency intersection occurs at $\epsilon_{r2}=0.6438$, which results in $f_c=5.958$GHz -- a slight increase as compared to the cutoff frequency of the homogeneously vacuum-filled waveguide. Since each of these intersections corresponds to a cutoff frequency of the  {\em HE}$_{11}$ mode, it is evident that the lined waveguide supports dual-band operation (this is verified by transmission simulations presented in the next section). These cutoff frequencies may be adjusted simply by varying the liner's permittivity profile and/or the waveguide's dimensions using (8a) and/or a figure equivalent to Fig.~\ref{fig2:subfig1}.

\begin{figure}[!t]
\label{fig3}
\centering
\subfigure[]{
\includegraphics[width=3.25in]{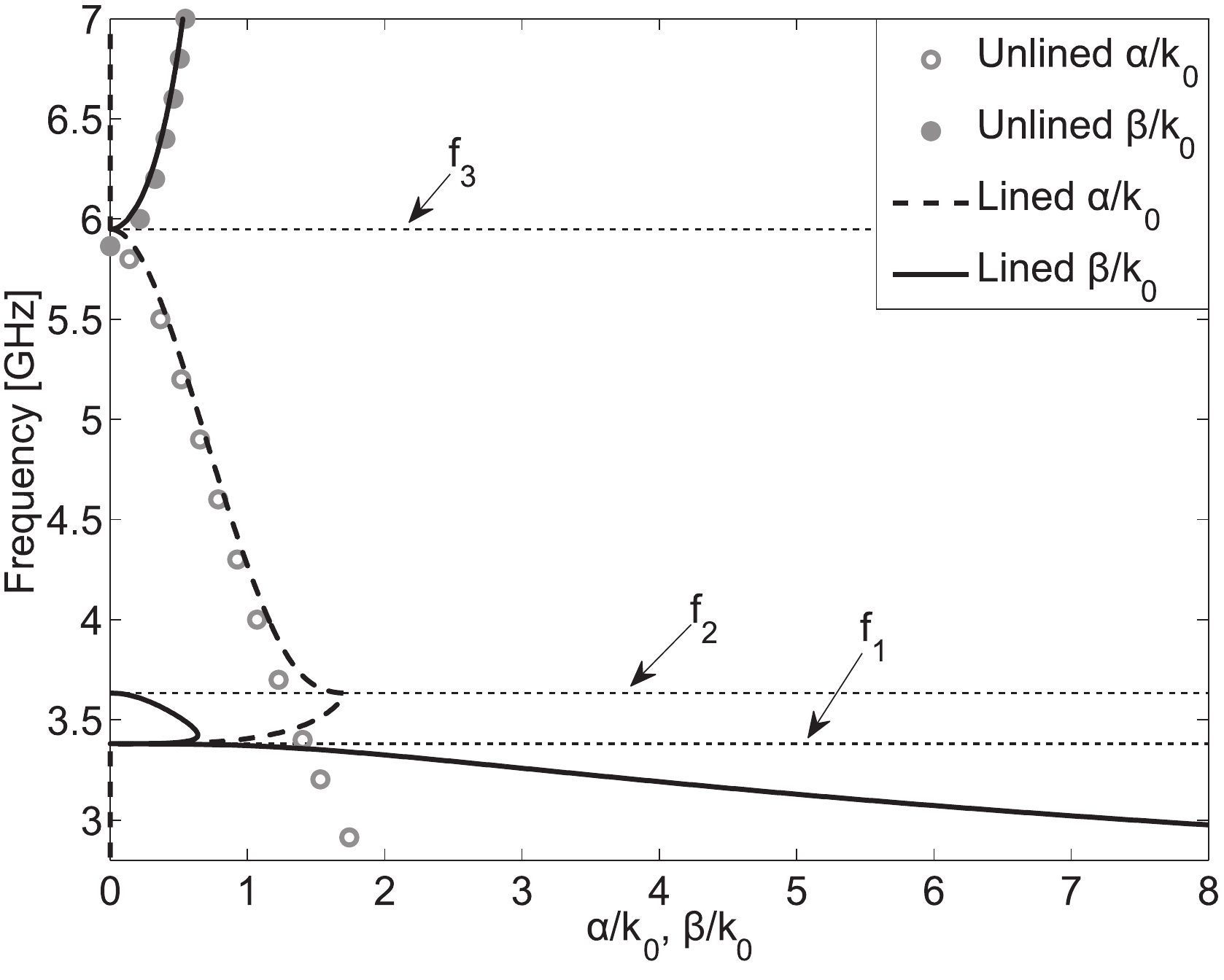}
\label{fig3:subfig1}
}
\subfigure[]{
\includegraphics[width=1.5in]{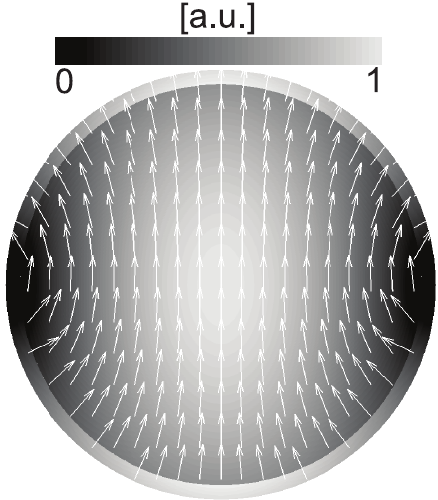}
\label{fig3:subfig2}
}
\subfigure[]{
\includegraphics[width=1.5in]{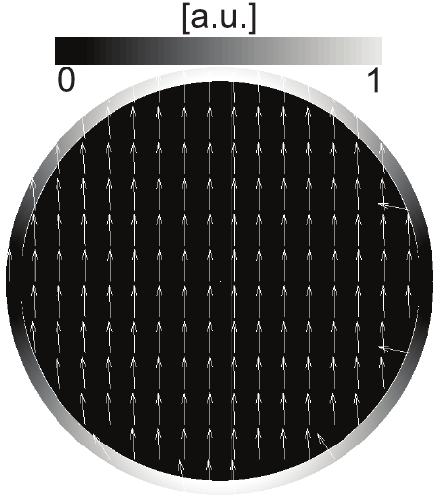}
\label{fig3:subfig3}
}
\caption{\unskip(a) Dispersion of $\alpha/k_0$ and $\beta/k_0$ for the metamaterial-lined waveguide's {\em HE}$_{11}$ mode (dashed and solid black lines, respectively) as compared to the dispersion of the {\em TE}$_{11}$ mode of an unlined waveguide of the same outer radius (empty and solid grey circles, respectively). The normalized complex electric-field magnitude and vectors in the transverse plane obtained from the theory in Sec.~\ref{sec:theory} for  the lined waveguide's at (b) $f_3=5.958$GHz and (c) $f_1=3.381$GHz, corresponding to the respective cutoff frequencies of the forward-wave and backward-wave bands.}
\end{figure}

Figure~\ref{fig3:subfig1} presents the {\em HE}$_{11}$-mode dispersion of the lossless metamaterial-lined waveguide as obtained from equations (7a)--(7c). Here, the solid and dashed lines, respectively, indicate $\beta/k_0$ and $\alpha/k_0$. For clarity, and consistent with the mode classification discussed in the previous section, a spectrum of strongly evanescent modes whose cutoffs lie below the frequency region shown have been omitted. Also shown for comparison is the {\em TE}$_{11}$-mode dispersion of a homogeneously vacuum-filled (unlined) waveguide of the same outer dimension, where $\beta/k_0$ and $\alpha/k_0$ are represented using grey solid and open circles, respectively. As suggested from the two intersections in Fig.~\ref{fig2:subfig1}, the metamaterial-lined waveguide supports two propagating bands: a forward-wave band near the cutoff of the unlined waveguide, and a frequency-reduced band well below this cutoff frequency. The frequency-reduced band has two regions of interest: a propagating backward-wave band for $f\leq f_1=3.381$GHz and a complex-propagation band for $f_1<f<f_2=3.644$GHz. Backward-wave propagation has previously been observed in inhomogeneous waveguides, albeit with conventional positive-valued dielectrics, and resulting in only a marginal reduction in cutoff frequency~\cite{clarricoatsBW}. The use of ENNZ materials in the present work suggests that backward-wave propagation can be achieved at dramatically reduced frequencies. As will be shown in the next section, the complex-propagation band does not allow for propagation of real power, a fact that may be attributed to the simultaneous existence of it's complex conjugate, which is excited equally in this region~\cite{chorney1961power}. The stopband in the region $f_2\leq f \leq f_3=5.958$GHz is explained by the fact that the metamaterial liner exhibits positive permittivities approaching unity as frequency is increased; as a result, propagation in the lined waveguide at these frequencies strongly resembles that in the unlined waveguide under its natural {\em TE}$_{11}$ cutoff. Accordingly, the forward-wave band in the region $f\geq f_3$ is also very similar in curvature and shape to that of the unlined waveguide. Furthermore, as evident from Fig.~\ref{fig2:subfig1}, liner permittivities in the EPNZ frequency region drive the waveguide even more deeply into cutoff. These results affirm that the metamaterial liner only significantly alters the propagation characteristics of the waveguide in the ENNZ frequency region.

The complex electric-field magnitudes, obtained using numerical implementation of the theory developed in Sec.~\ref{sec:theory}, are shown for both the lined waveguide's forward-wave band (Fig.~\ref{fig3:subfig2}) and its backward-wave band (Fig.~\ref{fig3:subfig3}), at their respective cutoff frequencies. As expected, the forward-wave {\em HE}$_{11}$-mode cutoff exhibits fields very similar to those at the {\em TE}$_{11}$-mode cutoff of an unlined waveguide. However, in the ENNZ regime, two major differences in the field profiles are observed: a decrease in field curvature in the inner vacuum region and a discontinuous higher concentration of fields in the metamaterial-liner region. The latter property can be explained using the normal electric-field boundary condition which requires that the normal component in the liner must increase as $|\epsilon_{r2}|\rightarrow 0$. It should be stressed that the field strengths in the vacuum region, although small in comparison to those in the liner, are not negligible. Furthermore, the reduction in frequency suggests a larger operating wavelength inside the vacuum region and, therefore, less field variation. It is interesting to note that as $\epsilon_{r2}\rightarrow 0^-$ the electric-field vectors are collimated, resembling those inside a parallel-plate waveguide. In fact, the fields in the liner assume the curvature needed to ensure that the tangential component of the electric field is zero at the outer PEC boundary.
\begin{figure}[!t]
\centering
\subfigure[]{
\includegraphics[width=3in]{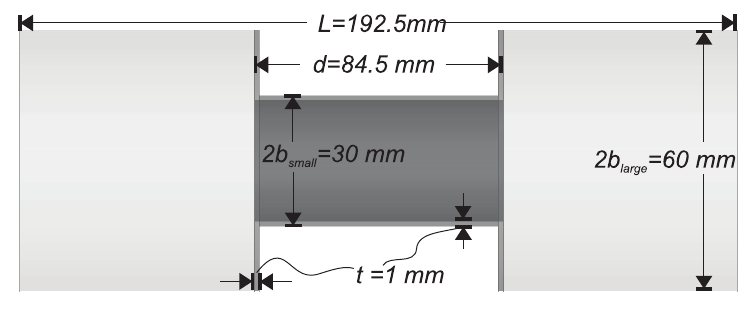}
\label{fig4:subfig1}
}
\subfigure[]{
\includegraphics[width=3in]{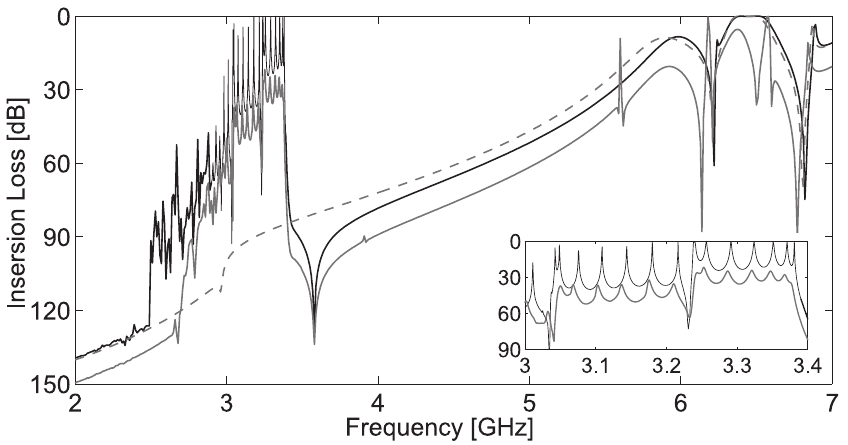}
\label{fig4:subfig2}
}
\caption{\unskip(a) Full-wave simulation model employed in the transmission analysis. A smaller below-cutoff waveguide is placed between two larger above-cutoff waveguides and a metamaterial liner possessing the complex dispersive permittivity in Sec.~\ref{sec:design} is introduced into the smaller waveguide. (b) Insertion loss for the unlined case (dashed grey line), the lined case with no loss (solid black line) and the lined case with loss (solid grey line). The inset shows in detail several resonances in the frequency-reduced backward-wave passband.}
\end{figure}
\section{Full-Wave Simulations}
\label{sec:trans}

Even though the cutoff frequency of the {\em HE}$_{11}$ mode can be reduced below the unlined waveguide's natural cutoff, it hasn't yet been shown if such structures are efficient at transporting power. To do so, a transmission analysis is performed using the full-wave simulation software HFSS~\cite{hfss} on the simulation model shown in Fig. \ref{fig4:subfig1}. In this representative setup, two larger vacuum-filled circular waveguides with a radius $b_{large}=30$mm and a {\em TE}$_{11}$ cutoff frequency of $2.928$GHz are connected by a smaller vacuum-filled waveguide with a radius $b_{small}=15$mm and a {\em TE}$_{11}$ cutoff frequency of $5.857$GHz. A waveport located at the end of one of the larger waveguide sections excites the {\em TE}$_{11}$ mode at frequencies that lie in the propagating region of the larger waveguide, but which correspond to the natural evanescent region of the smaller waveguide. The dashed curve in Fig.~\ref{fig4:subfig2} presents the insertion loss obtained for this setup and verifies that the intermediate waveguide under cutoff strongly attenuates the {\em TE}$_{11}$ mode.

Now, a metamaterial liner of thickness $t=1$mm is introduced into the smaller waveguide and assigned the dispersive permittivity, $\epsilon_{r2}(\omega)$, reported in Sec.~\ref{sec:design}. According to the theoretical results of the previous section, transmission should occur below $3.381$GHz for the frequency-reduced backward-wave band and above $5.958$GHz in the upper band. Inside the frequency-reduced band, the metamaterial liner would effectively enable a cross-sectional-area reduction of the unlined circular waveguide by a minimum of 75\%. A $1$mm-thick ring with the same material properties as the liner is placed on the PEC wall at the transition between the two waveguide sections in order to aid in the coupling of modes between the differently sized waveguides~\cite{silveirinha2007theory}. To understand the impact of losses in the metamaterial liner on the ability of the lined waveguide to transport power, both lossy ($\omega_t=5$MHz) and lossless ($\omega_t=0$) cases are compared. Figure~\ref{fig4:subfig2} shows the insertion loss in the lossless (solid black line) and lossy (solid grey line) cases. 
As predicted by the theory, the metamaterial liner results in a new passband below $f=3.381$GHz, corresponding to the backward-wave cutoff ($f_1$) in Fig.~\ref{fig3:subfig1}. An upper forward-wave band is also observed; however, its cutoff frequency is more difficult to infer from the insertion loss due to the excitation of high-order modes in the larger waveguide. By virtue of the particular geometrical parameters chosen in this arrangement (enabling the goal of a 75\% cross-sectional-area reduction), the {\em HE}$_{11}$ mode in the smaller waveguide couples to the {\em TM}$_{11}$ mode in the larger output waveguide for frequencies above $6.09$GHz. Another simulation with a smaller output waveguide of $b_{large}=20$mm and associated {\em TM}$_{11}$ mode cutoff frequency of $9.15$GHz (not shown) exhibited the expected smooth high-pass response.
\begin{figure}
\centering
\subfigure[]{
\includegraphics[width=1.5in]{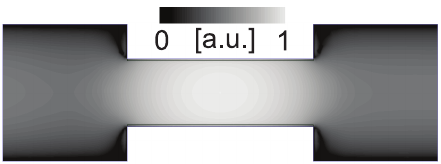}
\label{fig5:subfig1}
}
\subfigure[]{
\includegraphics[width=1.5in]{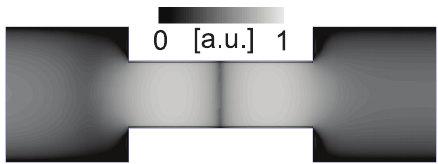}
\label{fig5:subfig2}
}
\subfigure[]{
\includegraphics[width=1.5in]{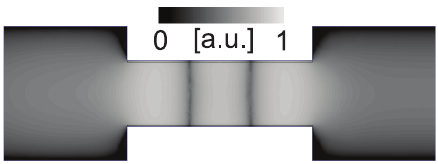}
\label{fig5:subfig3}
}
\subfigure[]{
\includegraphics[width=1.5in]{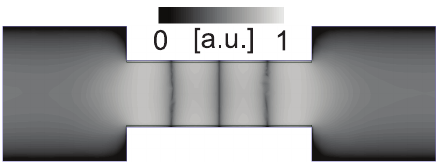}
\label{fig5:subfig4}
}
\subfigure[]{
\includegraphics[width=1.5in]{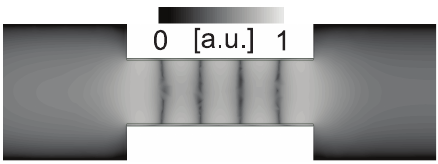}
\label{fig5:subfig5}
}
\subfigure[]{
\includegraphics[width=1.5in]{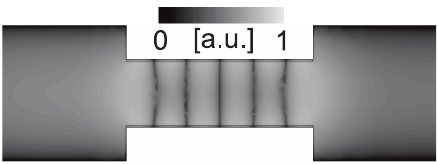}
\label{fig5:subfig6}
}
\caption{\unskip\unskip Complex electric-field magnitudes in the waveguides' H-plane at the following transmission peaks in the frequency-reduced backward-wave passband: (a) $f=3.381$GHz, (b) $f=3.371$GHz, (c) $f=3.351$GHz, (d) $f=3.324$GHz, (e) $f=3.291$GHz, and (f) $f=3.256$GHz. Each frequency corresponds to a resonant condition of an integer number of half-wavelengths supported by the lined waveguide section over its length.}
\end{figure}

The inset of Fig.~\ref{fig4:subfig2} presents the transmission features of the backward-wave region in greater detail, and reveals multiple narrow transmission peaks. In both the lossless and lossy case, a very fine frequency resolution is required to sample the maximum of each peak. When the liner is lossless, these peaks achieve total transmission of power through the lined waveguide section. Since the fields are concentrated strongly in the liner region, the introduction of loss degrades the transmission through the structure; nevertheless, the backward-wave band introduced by the liner still exhibits a dramatic increase in transmission over the unlined case, at times showing enhancements of over $56$dB. In fact, even in the lossy case, the transmission attains a peak value of $-21.7$dB at $f=3.253$GHz. Also evident in Fig.~\ref{fig4:subfig2} is a distinct antiresonance at $f=3.58$GHz which is located near the plasma frequency of the liner and the edge of the stop band (i.e., $f_2$ in Fig.~\ref{fig3:subfig1}), which appears not to be affected by the losses in the liner. The complex electric-field magnitude is plotted in the H-plane of the waveguide sections for the first six peaks in Fig.~\ref{fig5:subfig1}--\ref{fig5:subfig6}. The E-planes (not shown) exhibit very similar field distributions, except that the liner fields are more pronounced. Each successive peak corresponds to a resonant condition of an integer number of half-wavelengths supported by the lined waveguide section over its length. Moreover, that the order of the resonances increases as frequency decreases is characteristic of backward-wave propagation, and corroborates the dispersion data in Fig.~\ref{fig3:subfig1}. In all cases, it is evident that the incoming wavefront is restored at the outgoing side of the lined waveguide. These results validate that hollow waveguides lined using thin, ENNZ metamaterial liners support unusual resonant tunneling phenomena akin to the supercoupling observed in homogeneously ENZ-filled waveguides.

\begin{figure}
\centering
\subfigure[]{
\includegraphics[width=1.5in]{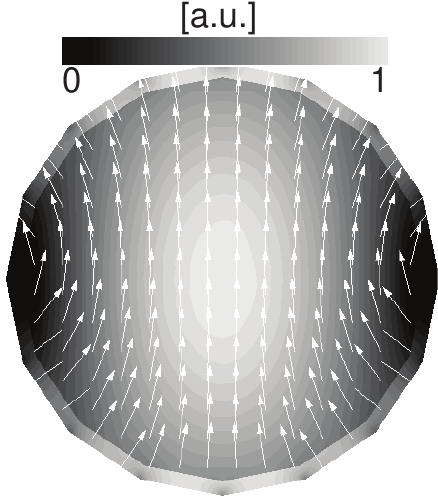}
\label{fig6:subfig1}
}
\subfigure[]{
\includegraphics[width=1.5in]{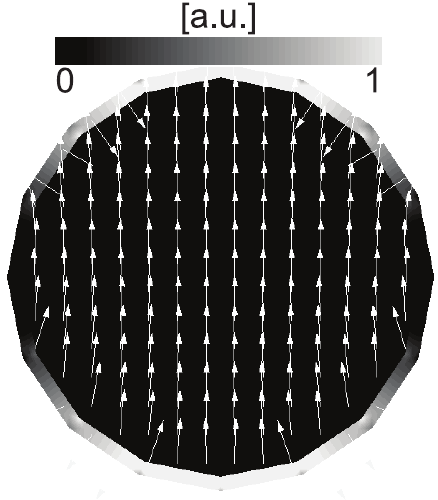}
\label{fig6:subfig2}
}
\caption{\unskip Simulated normalized electric field vectors and complex electric field magnitudes in the transverse cross-section at the center of the lined waveguide at (a) $f=5.958$GHz (b) $f=3.381$GHz. Close agreement is observed with the theoretical results of Figs.~\ref{fig3:subfig2} and \ref{fig3:subfig3}.}
\end{figure}

Figure~\ref{fig6:subfig1} presents the transverse electric field vectors in the lined waveguide's cross section at the cutoff frequency of the upper forward-wave band (i.e., $f=f_3=5.958$GHz). Superposed on these are the complex electric-field magnitudes which reveals a {\em TE}$_{11}$-like field distribution that matches well with that of Fig.~\ref{fig3:subfig2}. At the backward-wave band cutoff of $f=3.381$GHz where the first transmission peak is observed, Fig.~\ref{fig6:subfig2} verifies the results obtained from the theory (Fig.~\ref{fig3:subfig3}): that the transverse fields are strongly confined to the metamaterial liner and collimated in the vacuum region.

\begin{figure}
\centering
\includegraphics[width=3in]{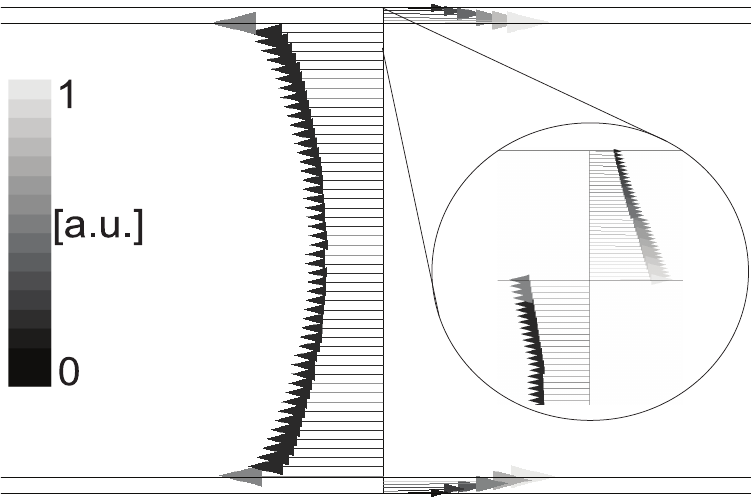}
{\caption{Transverse distribution of the Poynting-vector magnitude and direction at the center of the lined waveguide section at $3.381$GHz.}\label{fig7:subfig1}}
\end{figure}

It is well known that the direction of power flow can vary between different regions in inhomogeneous structures. For example, stacked DNG and DPS layers in a parallel-plate configuration can transport power forward in the DPS region that will, in turn, be coupled backward in the DNG region, resulting in zero net power being transported~\cite{alu2004guided}. An imbalance of power in each region can lead to non-zero net power being directed either forward or backward, as in the case of the conventional dielectric-lined waveguides~\cite{KWwhite}. Figure~\ref{fig7:subfig1} presents the simulated Poynting vector directions and magnitudes at the first resonant peak, along a diameter located at the center of the metamaterial-lined waveguide. The inset magnifies this behaviour at the interface between the vacuum and liner regions. It is evident that power is being transported in the positive direction through the liner and in the negative direction in the vacuum region at this particular location; however, it is also found (not shown) that the relative amounts of power in the liner and vacuum regions vary continuously along the guide. The total transmission of power at these resonant peaks, therefore, suggests a continuous resonant exchange of power between the liner and vacuum regions. It should be noted that this resonant transmission is achieved even though the liner occupies less than $13\%$ of the transverse cross-sectional surface area. This allows for the waveguide to be miniaturized while still leaving the majority of its volume empty for applications which require access to its interior.

\begin{figure}
\centering
\includegraphics[width=3in]{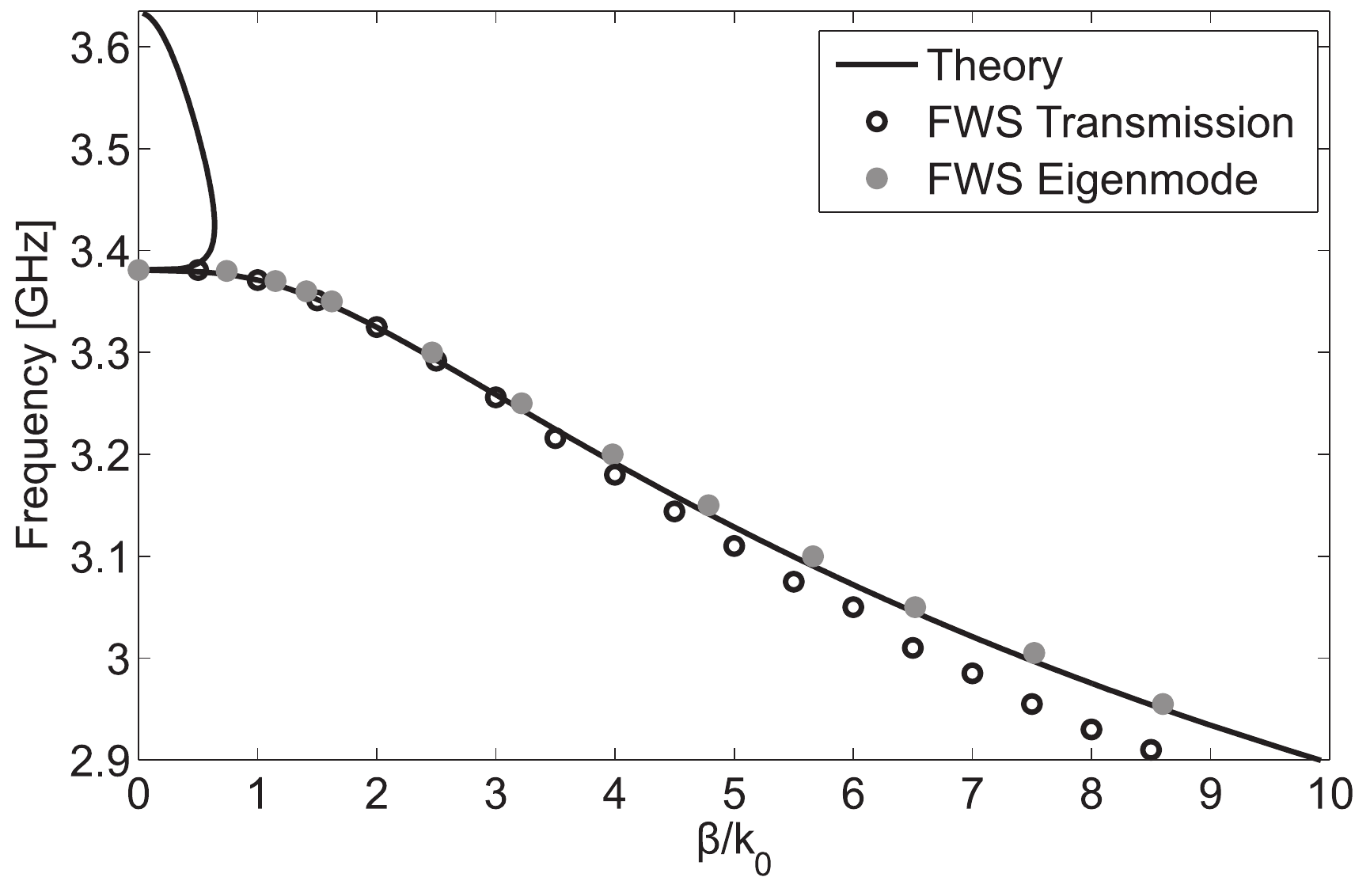}
{\caption{\unskip The dispersion curve of the frequency-reduced backward-wave passband obtained using three different methods: the full dispersion in equation (7a) (solid black line), full-wave eigenmode simulations (solid grey circles), and full-wave transmission simulations (empty black circles). }\label{fig8:subfig1}}
\end{figure}

To further validate the dispersion of the backward-wave band, Fig.~\ref{fig8:subfig1} compares the dispersion diagrams for the reduced {\em HE}$_{11}$ mode obtained using three different methods. The first method is based on the theoretically derived dispersion relation (7a) and is borrowed from Fig.~\ref{fig3:subfig1} (solid black line). The second method utilizes HFSS's eigenmode solver, but employs special techniques to overcome the following inherent limitations in the eigenmode solution process: first, only propagating modes (i.e. $\gamma\approx j\beta$) may be found; therefore, only the backward-wave and forward-wave (and not the complex) modes can be determined using this method. Second, the eigenmode solver determines the resonant frequencies of a particular geometry with known boundary conditions when the material properties at the solution frequency are known {\em a priori} (e.g., when the material properties are assumed constant with frequency). Consequently, in the case of dispersive materials, a conventional parametric sweep of phase shifts across the structure to obtain the dispersion curve is not a valid approach. However, knowing the frequency at which the dispersive metamaterial liner achieves a particular permittivity (by way of its known dispersion), the parametric sweep may be carried out using this fixed permittivity with the knowledge that the obtained dispersion curve is valid only at the corresponding frequency. By repeating this process over a number of fixed permittivity values corresponding to different frequencies, it is possible to obtain the true dispersion curve for the metamaterial-lined waveguide from HFSS's eigenmode solver. This is indicated in Fig.~\ref{fig8:subfig1} by the solid grey circles. The third method employs the full-wave transmission simulation results by examining the phase shift through the lined waveguide section at each resonant frequency (empty black circles). This process has been employed by other research groups with experimental results and has demonstrated relatively high precision, since all the peaks correspond to integer numbers of half-wavelength phase shifts~\cite{clarricoats1963circular}. The dispersion curves obtained from full-wave eigenmode simulations and theoretical analysis correlate strongly with one another across the whole backward-wave band, while the data derived from the transmission results deviates at higher values of $\beta/k_0$. This is to be expected, since we are comparing the dispersion of a finite-length lined-waveguide section to that of an infinitely long lined waveguide.

\section{Conclusion}

This work has presented a rigorous study of the propagation characteristics of metamaterial-lined circular waveguides. It was shown that the introduction of a thin liner possessing complex and dispersive permittivity into the waveguide volume results in a spectrum of hybrid modes, which may be classified as evanescent, propagating, or complex. Whereas the fundamental-mode cutoff is largely unaffected by the introduction of the liner, a frequency-reduced backward-wave passband appears well below cutoff when the real part of the liner permittivity assumes negative and near-zero values, and it is shown how the cutoff frequency of this mode may be designed using the dimensions and permittivities of the liner and waveguide. This mode also exhibits uniform, strongly collimated fields, with most of the field variation taking place in the liner. Full-wave transmission results through a lined waveguide section in the backward-wave region reveal multiple resonant transmission peaks corresponding to resonant phase conditions over the waveguide's length. Analytical and full-wave-simulation results of the dispersion properties and field profiles exhibit excellent agreement, and validate that metamaterial-lined circular waveguides operating in the frequency-reduced regime offer the potential for miniaturized waveguide components suitable for applications in which the waveguide volume must remain largely empty.
\section*{Acknowledgment}
The authors would like to thank Prof. Nader Engheta (University of Pennsylvania, Philadelphia, PA) for his encouragement and stimulating discussions.

\ifCLASSOPTIONcaptionsoff
  \newpage
\fi

\bibliography{bib}
\bibliographystyle{IEEEtran}

\end{document}